\documentclass[final, 12pt, oneside, print, peerreview, onecolumn]{ieeecolor}
\usepackage{tmi}
\usepackage{cite}
\usepackage{bm}
\usepackage{amsmath,amssymb,amsfonts}
\usepackage{algorithmic}
\usepackage{graphicx}
\usepackage{lineno}
\usepackage{subcaption}
\usepackage{here}
\usepackage{listings}

\newcommand{\argmax}{\mathop{\rm arg~max}\limits}

\usepackage{textcomp}
\def\BibTeX{{\rm B\kern-.05em{\sc i\kern-.025em b}\kern-.08em
    T\kern-.1667em\lower.7ex\hbox{E}\kern-.125emX}}
\renewcommand{\baselinestretch}{2.0}
\begin{document}
\title{X2CT-FLOW: Maximum a posteriori reconstruction using a progressive flow-based deep generative model for ultra sparse-view computed tomography in ultra low-dose protocols}
\author{Hisaichi Shibata
, Shouhei Hanaoka
, Yukihiro Nomura, \IEEEmembership{Member, IEEE}, Takahiro~Nakao
, Tomomi Takenaga,\\
Naoto Hayashi
, and Osamu Abe
\thanks{The Department of Computational Diagnostic Radiology and Preventive Medicine, The University of Tokyo Hospital, is sponsored by HIMEDIC Inc. and Siemens Healthcare K.K.
This work was supported by JSPS KAKENHI Grant Number 21K18073.}
\thanks{H. Shibata, Y. Nomura, T. Nakao, T. Takenaga and N. Hayashi are with the Department of Computational Diagnostic Radiology and Preventive Medicine, The University of Tokyo Hospital, 7-3-1 Hongo, Bunkyo-ku, Tokyo 113-8655, Japan (e-mail sh@g.ecc.u-tokyo.ac.jp, nomuray-tky@umin.ac.jp, tanakao-tky@umin.ac.jp, takenaga-tky@umin.ac.jp, naoto-tky@umin.ac.jp). }
\thanks{S. Hanaoka and O. Abe are with the Department of Radiology, The University of Tokyo Hospital, 7-3-1 Hongo, Bunkyo-ku, Tokyo 113-8655, Japan (e-mail hanaokalog@gmail.com, abediag@g.ecc.u-tokyo.ac.jp).}
\thanks{Y. Nomura is with the Center for Frontier Medical Engineering, Chiba University, 1-33 Yayoi-cho, Inage-ku, Chiba 263-8522, Japan.}
\thanks{O. Abe is with the Division of Radiology and Biomedical Engineering, Graduate School of Medicine, The University of Tokyo, 7-3-1 Hongo, Bunkyo-ku, Tokyo 113-8656, Japan.}
}

\maketitle

\begin{abstract}
Ultra sparse-view computed tomography (CT) algorithms can reduce radiation exposure of patients, but those algorithms lack an explicit cycle consistency loss minimization and an explicit log-likelihood maximization in testing.
Here, we propose X2CT-FLOW for the maximum a posteriori (MAP) reconstruction of a three-dimensional (3D) chest CT image from a single or a few two-dimensional (2D) projection images using a progressive flow-based deep generative model, especially for ultra low-dose protocols.
The MAP reconstruction can simultaneously optimize the cycle consistency loss and the log-likelihood.
The proposed algorithm is built upon a newly developed progressive flow-based deep generative model, which is featured with exact log-likelihood estimation, efficient sampling, and progressive learning.
We applied X2CT-FLOW to reconstruction of 3D chest CT images from biplanar projection images without noise contamination (assuming a standard-dose protocol) and with strong noise contamination (assuming an ultra low-dose protocol).
With the standard-dose protocol, our images reconstructed from 2D projected images and 3D ground-truth CT images showed good agreement in terms of structural similarity (SSIM, 0.7675 on average), peak signal-to-noise ratio (PSNR, 25.89 dB on average), mean absolute error (MAE, 0.02364 on average), and normalized root mean square error (NRMSE, 0.05731 on average).
Moreover, with the ultra low-dose protocol, our images reconstructed from 2D projected images and the 3D ground-truth CT images also showed good agreement in terms of SSIM (0.7008 on average), PSNR (23.58 dB on average), MAE (0.02991 on average), and NRMSE (0.07349 on average).
\end{abstract}


\section{Introduction}
\label{sec:Introduction}
\IEEEPARstart{X}{-ray} chest computed tomography (CT) is a three-dimensional (3D) image modality. It has diagnostic superiority over chest X-rays (CXRs), but patients have greater radiation exposure than in the case of CXRs \cite{nam2021image}.
To reduce radiation exposure, sparse-view CTs have been developed.
Typical sparse-view CTs adopt a \textit{maximum a posteriori} (MAP) reconstruction, which can reduce the number of projection images for CT reconstruction.
Those sparse-view CTs adopt a prior that assumes a sparsity of images, e.g., regularization terms of quadratic form in \cite{levitan1987maximum} and the $l_1$ norm in compressed sensing \cite{baraniuk2007compressive}.
Sparse-view CTs are used to reconstruct a 3D image from tens of two-dimensional (2D) projection images, but Shen and coworkers \cite{shen2021geometry, shen2019patient} proposed ultra sparse-view CT algorithms to reconstruct a 3D image from a single or a few projection images.
A similar work by Ying et al. \cite{ying2019x2ct} reconstructed a 3D CT image from biplanar CXR images.
However, previous algorithms related to ultra sparse-view CT \cite{shen2021geometry,shen2019patient, ying2019x2ct,peng2020xraysyn,henzler2018single} adopt end-to-end supervised deep neural networks without exception: those algorithms do not handle MAP reconstruction, in which log-likelihood and cycle consistency loss are simultaneously optimized.
The lack of the optimization of log-likelihood means that there is no explicit guarantee that those algorithms can reconstruct images that are likely to be the 3D ground-truth CT images.
The lack of the optimization of the cycle consistency loss means that there is no explicit guarantee that the reconstructed 3D image projected onto a 2D plane coincides with the input 2D projection image.
These missing factors can potentially deprive these ultra sparse-view CT algorithms of robustness against noise.
The lack of robustness is especially problematic in ultra low-dose protocols, where strong noise significantly contaminates the 2D projection images.

\begin{figure}
    \centering
    \includegraphics[width=8cm]{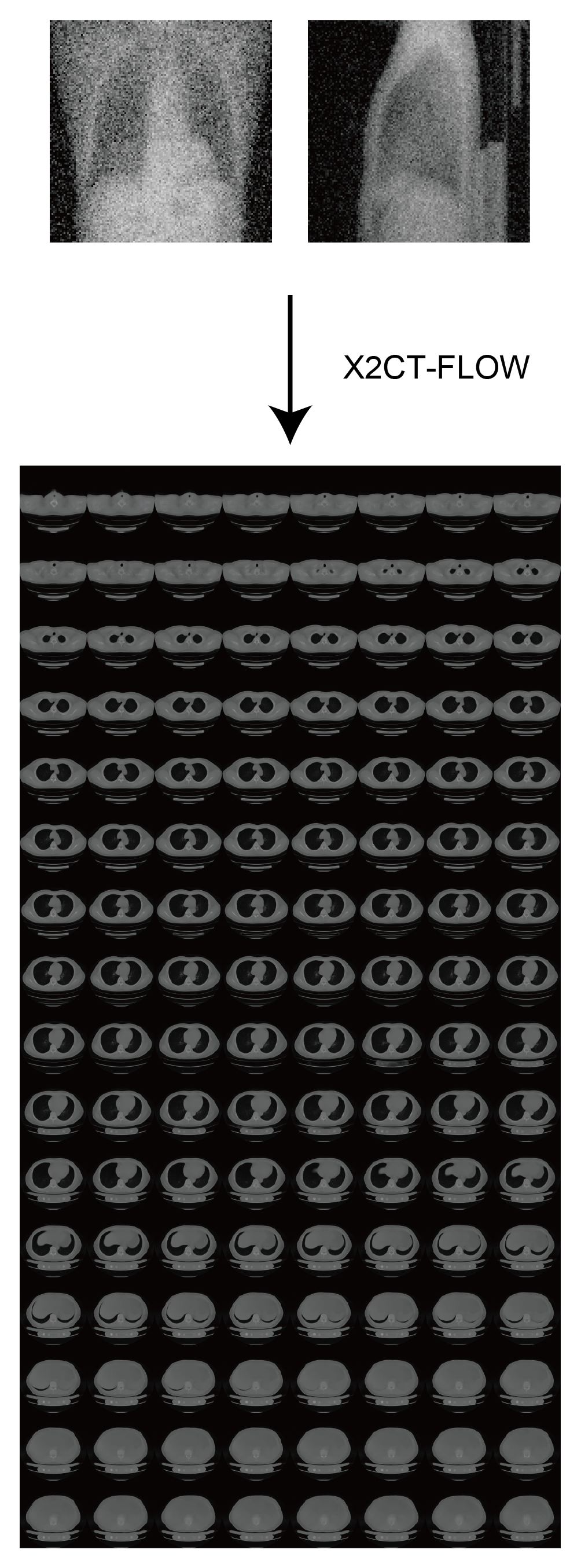}
    \caption{X2CT-FLOW can find the optimum 3D chest CT image (the bottom) from a single or a few noisy projection images (the top) with MAP reconstruction. Scales of the top and bottom images are not the same.}
    \label{fig:x2ct_flow}
\end{figure}

Here, we propose a novel ultra sparse-view algorithm especially for ultra low-dose protocols (\textbf{X2CT-FLOW}, Fig.~\ref{fig:x2ct_flow}), which adopts the MAP reconstruction.
Unlike ordinal compressed sensing, we do not explicitly impose sparsity on reconstructed images for a prior with the regularization terms; instead, we train the prior with a progressive flow-based deep generative model with 3D chest CT images.
The MAP reconstruction can simultaneously optimize the log-likelihood and the cycle consistency loss of a reconstructed image in testing (for details, see Sec.~\ref{sec:methods}).
We build the proposed algorithm on 3D GLOW \cite{shibata2021}, which is one of the flow-based deep generative models; the models can execute exact log-likelihood estimation and efficient sampling \cite{kobyzev2020normalizing}.
Furthermore, we realize training with high-resolution ($128^3$) 3D chest CT images with progressively increasing image gradations (\textbf{progressive learning}), and showcase a high-resolution 3D model.
To the best of our knowledge, there is no previous study of the flow-based generative models in which such a high-resolution model was showcased.

In summary, the contributions of this paper are as follows:
\begin{enumerate}
    \item We propose the MAP reconstruction for ultra sparse-view CTs, especially for ultra low-dose protocols, and validate it using digitally reconstructed radiographs.
    \item We adopt an exact log-likelihood of 3D chest CT images estimated with a 3D flow-based deep generative model to optimize the prior.
    \item We adopt an invertible decoder of 3D flow-based deep generative model to optimize the cycle consistency loss.
    \item We establish progressive learning to realize high-resolution 3D flow-based deep generative models.
    \item We showcase a 3D flow-based deep generative model of 3D chest CT images, which has state-of-the-art resolution ($128^3$).
\end{enumerate}

\section{Methods}
\label{sec:methods}
We introduce a 2D projection image vector $\bm{y}_i^j$ whose dimensions are $H_{2D} \times W_{2D} \times C_{2D}$ and a 3D chest CT image vector $\bm{x}_i$ whose dimensions are $D_{3D}\times H_{3D}\times W_{3D} \times C_{3D}$, where $H_{2D}$, $W_{2D}$, and $C_{2D}$ are the height, width, and channel size of the 2D image and $D_{3D}$, $H_{3D}$, $W_{3D}$, and $C_{3D}$ are the depth, height, width, and channel size of the 3D image, respectively.
The subscript $i$ distinguishes patients and we omit it if not necessary, and the superscript $j$ distinguishes different view angle images for each patient, where $1\le j\le N$ and $N$ is the number of the angles, e.g., $N=1$ for a uniplanar (single) image and $N=2$ for biplanar images.
To simplify the explanation below, we set $N=1$; hence, we omit the superscript $j$.
We show formulations in cases of $N\ge 2$ in Appendix~\ref{app:formulation_n_gt_1}.
We first train a flow-based deep generative model (3D GLOW) using a set of 3D chest CT images, and then reconstruct a 3D chest CT image from a single or a few 2D projection images (X2CT-FLOW).

\subsection{3D GLOW}
\begin{figure}
    \centering
    \includegraphics[width=5cm]{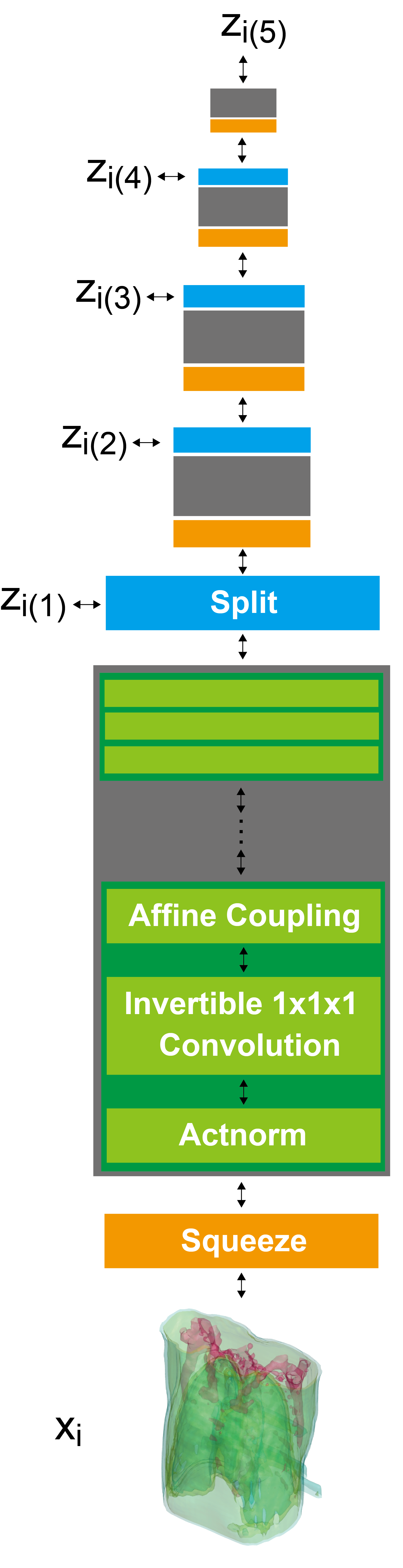}
    \caption{Deep neural network architecture of 3D GLOW. $\bm{x}_i$ represents a 3D CT image vector and ${\bm{z}}_{i(k)}, k=1, \ldots, 5$ represent the latent variable vectors in each deep neural network level. We rendered the 3D chest CT image with three iso-surfaces.}
    \label{fig:3d_glow}
\end{figure}

In training, the flow-based deep generative models minimize the Kullback–Leibler divergence between the true distribution $\left[p(\bm{x}_i)\right]$ and the estimated distribution $\left[p_{\bm{\theta}}(\bm{x}_i)\right]$ of input images (i.e., 3D chest CT images) by minimizing the negative log-likelihood (NLL) as,
\begin{eqnarray}
    \mathcal{L}\left( \mathcal{D}\right) &=& - \frac{1}{|\mathcal{D}|} \sum_{\bm{x}_i \in \mathcal{D}} \log{p_{\bm{\theta}}\left( \bm{x}_i \right)},
\end{eqnarray}
where the subscript $\bm{\theta}$ represents parameters in the model, $\mathcal{D}$ represents a set of images for training, $|\mathcal{D}|$ is the number of images for the training, and the subscript $i$ distinguishes each image.
The NLL is not tractable; therefore, we map the NLL onto a tractable simpler distribution (e.g., a multivariate independent normal distribution) as:
\begin{eqnarray}
    \log{p_{\bm{\theta}} (\bm{x}_i)} = \log{p \left( \bm{z}_i \right)} - \log \left|\det \left(\frac{\partial \bm{G}_{\bm{\theta}}}{\partial \bm{z}_i}\right)\right|,
\end{eqnarray}
where $p\left( \bm{z}_i\right)$ is the tractable probability density function, e.g., the standard normal distribution $\bm{z}_i \sim \mathcal{N}(\bm{0},\bm{I})$, and $\bm{x}_i = \bm{G}_{\bm{\theta}}(\bm{z}_i)$ is the invertible decoder in the model.
We adopt 3D GLOW \cite{shibata2021}, which is a 3D extension of one of the state-of-the-art 2D flow-based deep generative models, GLOW \cite{kingma2018glow}.
We indicate the concrete form of $\bm{G}_{\bm{\theta}}$, i.e., the deep neural network architecture of 3D GLOW, in Fig.~\ref{fig:3d_glow}.

Here, for the first time, we propose to train the flow-based deep generative models in a progressive manner to accelerate the convergence of the NLL.
Firstly, we train 3D GLOW with 2 bits images and then 3 bits, 4 bits, and finally 8 bits images.
We show codes to reproduce the reduction of image gradation in Appendix~\ref{app:codes_progressive_learning}.
Moreover, we show the beneficial effects of the progressive learning in Appendix~\ref{app:progressive_learning}.

By using a trained 3D GLOW model, we can generate fictional but realistic images, i.e., sampling, as follows:
\begin{eqnarray}
    \bm{z}_i &\sim& \mathcal{N}\left( \bm{\mu}_{\bm{\theta}}, T^2 \cdot \Sigma^2_{\bm{\theta}}\right), \\
    \bm{x}_i &=& \bm{G}_{\bm{\theta}} \left( \bm{z}_i \right),
\end{eqnarray}
where $T$ (scalar) is the temperature for the reduced-temperature model \cite{parmar2018image}, $\bm{\mu}_{\bm{\theta}}$ is the estimated means of the images for training in the latent space, and $\Sigma^2_{\bm{\theta}}$ (diagonal matrix) is the estimated variances of the images for training in the latent space.
For details of the flow-based deep generative models, see \cite{dinh2014nice, dinh2016density, kingma2018glow}.

\subsection{X2CT-FLOW}
In testing, we reconstruct the 3D image from a single or a few noisy 2D projection images.
We define a linear observation matrix $P$ as follows:
\begin{eqnarray}
    \bm{y}_{h,w,c} &=& \left( P \bm{x}\right)_{h,w,c} \\
    &\equiv& \frac{1}{D_{3D}} \sum^{D_{3D}}_{d=1} x_{d,h,w,c},
\end{eqnarray}
where the indices $d, h, w$, and $c$ distinguish voxels and the observation matrix $P$ is a linear operator to average voxels in the depth direction.
We can similarly define the observation matrices for different projection directions.
First, we adopt the matrix to emulate 2D projection images $\bm{y}$ obtained with an ultra sparse-view CT from an image $\bm{x}$ obtained with a standard CT, i.e., forward projection.
In this study, we do not use 2D projection images obtained with an ultra sparse-view CT.
Second, we adopt the matrix to reconstruct $\bm{x}$ from $\bm{y}$, i.e., back projection.
We find $\hat{\bm{x}}$ such that it maximizes the log-posterior of $\bm{x}$ given the observation fact $\bm{y}$, i.e., $\log p(\bm{x}|\bm{y})$.
We created $\bm{y}$ so that the probabilistic distribution of noise on $\bm{y}$ follows a normal distribution $\left[ \mathcal{N}(\bm{0}, \sigma^2\bm{I}) \right]$.
Therefore, we have
\begin{eqnarray}
    \label{eqn:observation_eq}
    \bm{y} &=& P \bm{x} + \sqrt{\sigma^2} \bm{w}, \\
    \bm{w} &\sim& \mathcal{N}(\bm{0}, \bm{I}), 
\end{eqnarray}
where $\sigma^2$ is the variance of the normal noise (scalar)  and $\bm{w}$ is a normal noise vector.
Equation~(\ref{eqn:observation_eq}) means that $\log p (\bm{y} | \bm{x})$ follows a normal distribution for fixed $\bm{x}$ and $P$.
By using the above definitions, we finally have
\begin{eqnarray}
 \label{eqn:bayes}
    \hat{\bm{x}} &=& \argmax_{\bm{x}} \log p (\bm{x} | \bm{y}) \nonumber \\
    &=& \argmax_{\bm{x}} \log p (\bm{y} | \bm{x})  + \log p(\bm{x}) - \log p (\bm{y}) \nonumber\\
    &=& \argmax_{\bm{x}} \log p (\bm{y} | \bm{x})  + \log p(\bm{x}) \nonumber\\
    &=&  \argmax_{\bm{x}} \log \left[ \frac{1}{\sqrt{2\pi \sigma^2}}\exp{\left(-\frac{1}{2} \bm{w}^T \bm{w}\right)}  \right] + \log p(\bm{x})  \nonumber \\
    &=& \argmax_{\bm{x}} -\frac{1}{2} \log{2\pi \sigma^2} - \frac{1}{2 \sigma^2} \|  \bm{y} - P \bm{x} \|_2^2 + \log p(\bm{x}) \nonumber\\
    &=& \argmax_{\bm{x}} -\frac{1}{2 \sigma^2} \|  \bm{y} - P \bm{x} \|_2^2 + \log p(\bm{x})  \\
    &\equiv& \argmax_{\bm{x}} - \mathcal{E}\left(\bm{x}\right),
\end{eqnarray}
where between the first and the second lines, we applied Bayes' theorem.
The first term of Eq.~(\ref{eqn:bayes}) is the cycle consistency loss and the second term of Eq.~(\ref{eqn:bayes}) is the log-likelihood term.
We approximate the log-likelihood term $\left[ \log{p (\bm{x})} \right]$ by $\left[ \log{p_{\bm{\theta}} (\bm{x})} \right]$ using a trained 3D GLOW model.

On the basis of Eq.~(\ref{eqn:bayes}), we reconstruct the optimum 3D chest CT image from each chest 2D projection image in a testing dataset.
We adopt the gradient descent method to obtain $\hat{\bm{x}_i}$ such that it can satisfy Eq.~(\ref{eqn:bayes}), i.e.,
\begin{eqnarray}
\label{eqn:gdm_x}
    \bm{x}^{(n+1)}_i \leftarrow \bm{x}^{(n)}_i - \alpha \cdot \nabla_{\bm{x}_i} \mathcal{E}\left( \bm{x}_i^{(n)} \right),
\end{eqnarray}
where $\alpha$ is an empirical relaxation coefficient and the superscript $n$ is an iteration number.
Furthermore, to accelerate the convergence of Eq.~(\ref{eqn:gdm_x}), we adopt an invertible decoder $\bm{G}_{\bm{\theta}}$ of 3D GLOW, which can map a latent vector $\bm{z}_i$ to a 3D chest CT image $\bm{x}_i$, i.e., $\bm{x}_i = \bm{G}_{\bm{\theta}}(\bm{z}_i)$.
Finally, we adopt the gradient descent method to obtain $\hat{\bm{z}_i}$ such that $\hat{\bm{z}_i}$ can satisfy Eq.~(\ref{eqn:bayes}), i.e.,
\begin{eqnarray}
\label{eqn:gdm_z}
    \bm{z}^{(n+1)}_i \leftarrow \bm{z}^{(n)}_i - \alpha \cdot \nabla_{\bm{z}_i} \mathcal{E}\left[ \bm{G}_{\bm{\theta}}(\bm{z}_i^{(n)} )\right],
\end{eqnarray}
and if the $l_2$ norm between the current latent vector $\bm{z}_i^{(n+1)}$ and the previous latent vector $\bm{z}_i^{(n)}$ converges, we can obtain the optimum 3D chest CT image $\hat{\bm{x}}_i$ as
\begin{eqnarray}
    \hat{\bm{x}}_i = \bm{G}_{\bm{\theta}}( \hat{\bm{z}}_i).
\end{eqnarray}

\section{Numerical Experiment}
\label{sec:numerical_experiment}

\subsection{Materials}
This study was approved by the ethical review board of our institution, and written informed consent to use the images was obtained from all the subjects.
We used chest CT images of 450 normal subjects.
These images were scanned at our institution with a GE LightSpeed CT scanner (GE Healthcare, Waukesha, WI, USA).
The acquisition parameters were as follows: number of detector rows, 16; tube voltage, 120 kVp; tube current, 50--290 mA (automatic exposure control); noise index, 20.41; rotation time, 0.5 s; moving table speed, 70 mm/s; body filter, standard; reconstruction slice thickness and interval, 1.25 mm; field of view, 400 mm; matrix size, 512$\times$512 pixels; pixel spacing, 0.781 mm.
We randomly divided the images of the 450 normal subjects into training (384), validation (32), and test datasets (34).
We converted the acquired images $I_\mathrm{src}$ (CT number in HU units) into images $I_\mathrm{dst}$ with the following empirical formula:
\begin{eqnarray}
    I_\mathrm{dst} = \frac{255\cdot\left\{  \mathrm{clip}\left[ I_\mathrm{src}, -1000, \max\left( I_\mathrm{src}\right) \right] + 1000\right\}}{\max\left( I_\mathrm{src}\right) + 1000}, 
\end{eqnarray}
where the operator $\mathrm{clip}(x, a, b)$ restricts the value range of an array $x$ from $a$ to $b$, and the operator $\mathrm{max}(x)$ returns the maximum value in $x$.
Owing to limits in GPU memory, we down-sampled $I_{\mathrm{dst}}$ to the resolution of $128^3$; hence, we set $D_{3D}=H_{3D}=W_{3D}=H_{2D}=W_{2D}=128$ and $C_{3D}=C_{2D}=1$.

\subsection{3D GLOW}
X2CT-FLOW is built upon 3D GLOW.
To enhance the stability of the training of 3D GLOW, we modified the scale function in the affine coupling layer to the scale $s\left( h_2 + 2.0\right) + \epsilon$ from the scale $s\left( h_2 + 2.0\right)$, where $s$ is the sigmoid function, $h_2$ is the input from the previous split layer, and $\epsilon$ is a newly introduced hyperparameter.
We empirically set $\epsilon=10^{-3}$.

\begin{table}[htb]
  \begin{center}
    \caption{Hyperparameters used to train 3D GLOW model.}
    \label{tab:hps}    
    \begin{tabular}{lc} \hline
      Flow coupling & Affine \\
      Learn-top option & True \\
      Flow permutation & 1$\times$1$\times$1 convolution \\
      Minibatch size & 1 per GPU \\
      Train epochs & 96 (2 bits) \\
      & 324 (3 bits from 2 bits) \\
      & 24 (4 bits from 3 bits) \\ 
      & 144 (8 bits from 4 bits) \\      
      Layer levels & 5 \\
      Depth per level & 8 \\
      Filter width & 512 \\
      Learning rate in steady state & $1.0 \times 10^{-4}$ \\
      \hline
    \end{tabular}
    \end{center}
\end{table}
The hyperparameters used to train the model are listed in Table~\ref{tab:hps}.
We utilized Tensorflow 1.14.0 for the back end of the DNNs.
The CUDA and cuDNN versions used were 10.0.130 and 7.4, respectively.
All processes were carried out on a workstation consisting of two Intel Xeon Gold 6230 processors, 384 GB memory, and five GPUs (NVIDIA Quadro RTX 8000 with 48 GB memory).
For the training, we used only four GPUs out of the five GPUs, and for the testing, we utilized only one GPU.

\section{Results}
\label{sec:results}

\subsection{Standard-dose protocol}
\begin{figure}[t]
  \begin{minipage}[b]{0.45\hsize}
    \centering
    \includegraphics{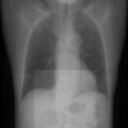}
    \subcaption{}\label{fig:input1of2}
  \end{minipage}
  \begin{minipage}[b]{0.45\hsize}
    \centering
    \includegraphics{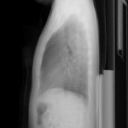}
    \subcaption{}\label{fig:input2of2}
  \end{minipage}\\
  \begin{minipage}[b]{0.45\hsize}
    \centering
    \includegraphics{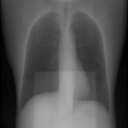}
    \subcaption{}\label{fig:initial1of2}
  \end{minipage}
  \begin{minipage}[b]{0.45\hsize}
    \centering
    \includegraphics{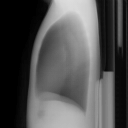}
    \subcaption{}\label{fig:initial2of2}
  \end{minipage}\\
  \begin{minipage}[b]{0.45\hsize}
    \centering
    \includegraphics{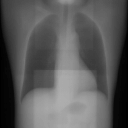}
    \subcaption{}\label{fig:conv1of2}
  \end{minipage}
  \begin{minipage}[b]{0.45\hsize}
    \centering
    \includegraphics{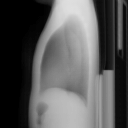}
    \subcaption{}\label{fig:conv2of2}
  \end{minipage}    
  \caption{Input images and projections of a reconstructed image ($N=2$): (a, b) input images, (c, d) projections of an initial guess image (sampled with temperature $T=0.5$), (e, f) projections of the optimum reconstructed image. The intensities of these images have been modified to enhance visibility.}
  \label{fig:inout}
\end{figure}

\begin{figure}[htbp]
    \centering
    \includegraphics[width=8.5cm]{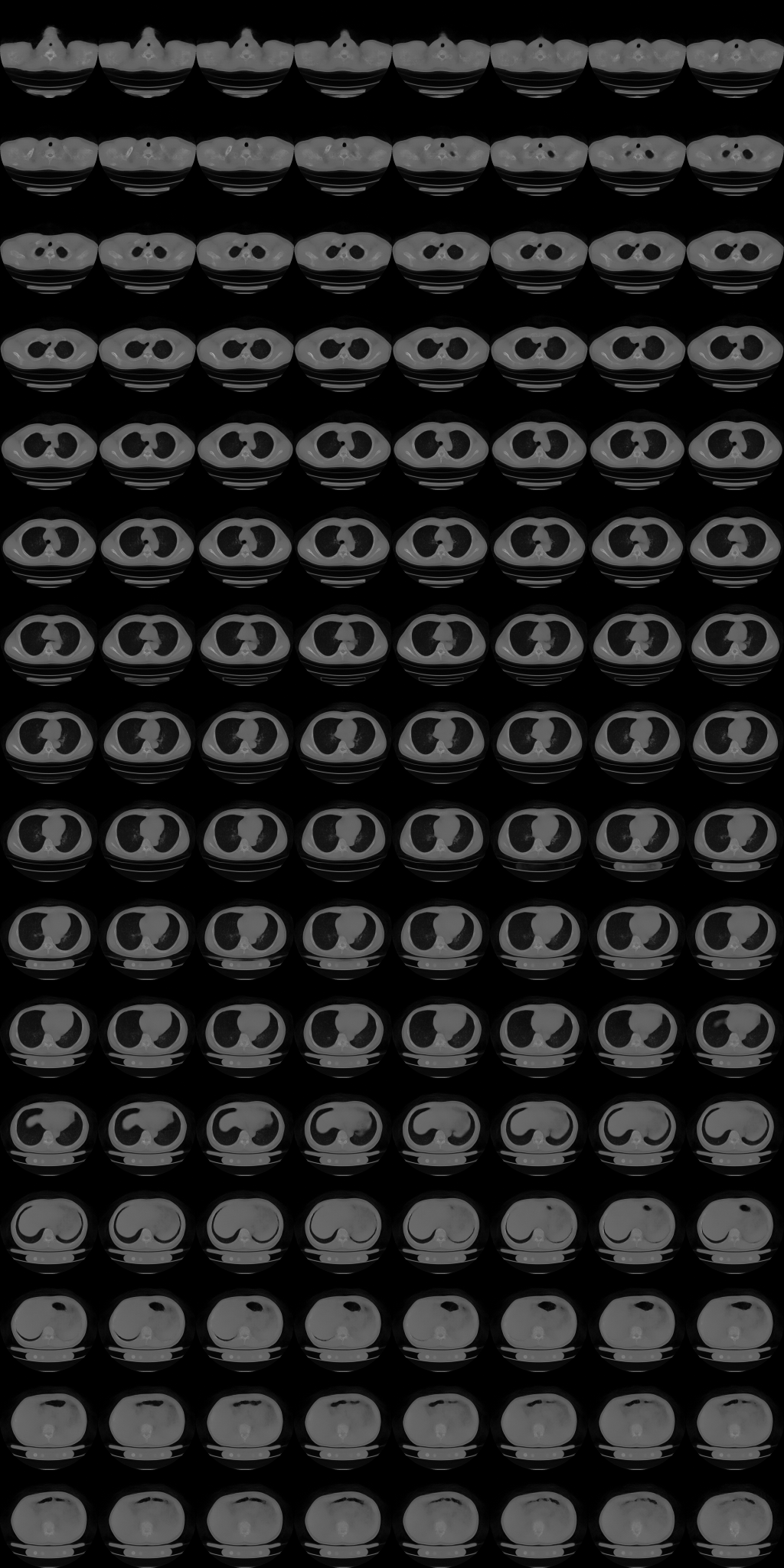}
    \caption{Reconstructed 3D CT image with X2CT-FLOW from Figs.~\ref{fig:inout}(a) and (b) ($\sigma^2=0, N=2$).}
    \label{fig:typical_reconstructed_image_2_planes_wo_noise}
\end{figure}

\begin{figure}[htbp]
    \centering
    \includegraphics[width=8.5cm]{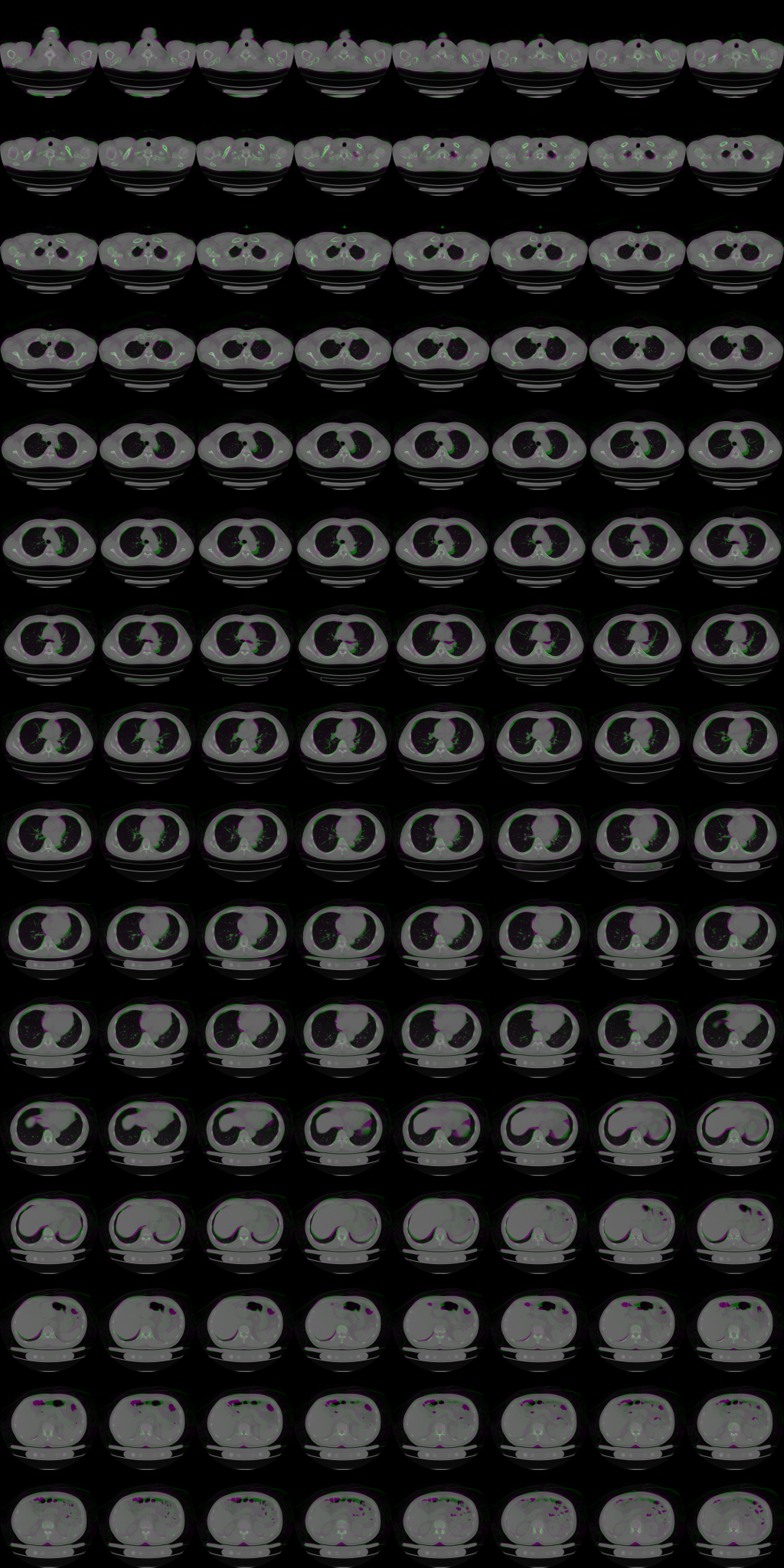}
    \caption{Superposition of the reconstructed 3D CT image shown in Fig.~$\ref{fig:typical_reconstructed_image_2_planes_wo_noise}$ (magenta) and the ground-truth image (green).}
    \label{fig:typical_diff_reconstructed_image_2_planes_wo_noise}
\end{figure}

\begin{figure}[t]
  \begin{minipage}[b]{0.45\hsize}
    \centering
    \includegraphics{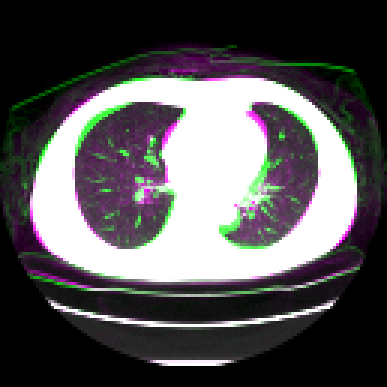}
    \subcaption{}\label{fig:enlarge_33_axial}
  \end{minipage}
  \begin{minipage}[b]{0.45\hsize}
    \centering
    \includegraphics{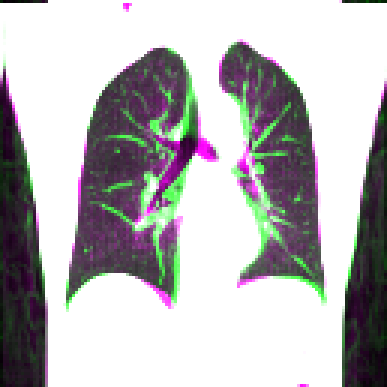}
    \subcaption{}\label{fig:enlarge_33_coronal}
  \end{minipage}\\
  \begin{minipage}[b]{0.45\hsize}
    \centering
    \includegraphics{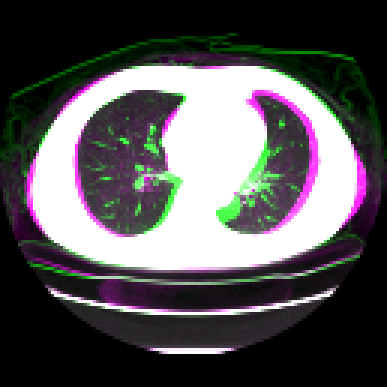}
    \subcaption{}\label{fig:enlarge_33_axial_2}
  \end{minipage}
  \begin{minipage}[b]{0.45\hsize}
    \centering
    \includegraphics{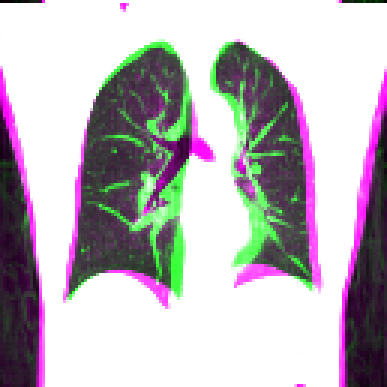}
    \subcaption{}\label{fig:enlarge_33_coronal_2}
  \end{minipage} 
    \caption{Superposition of the reconstructed 3D CT image (magenta) and the ground-truth image (green), in pulmonary window setting. (a, b) Partially enlarged axial and coronal views of Fig.~\ref{fig:typical_diff_reconstructed_image_2_planes_wo_noise}. (c, d) Partially enlarged axial and coronal views of Fig.~\ref{fig:typical_diff_reconstructed_image_2_planes_with_noise}.}
  \label{fig:enlarge_lung}
\end{figure}

We assume the limit of $\sigma^2 \rightarrow 0$.
In this limit, we have
\begin{eqnarray}
\mathcal{E}\left[\bm{G}_{\bm{\theta}}\left(\bm{z}\right) \right] \rightarrow \frac{1}{2 \sigma^2} \|  \bm{y} - P \bm{G}_{\bm{\theta}}\left( \bm{z} \right) \|_2^2.
\end{eqnarray}
We put $\alpha = 0.2 \sigma^2 \cdot \left[ 1 - \exp{(-0.01 \cdot n)} \right]$ and iterated while $n\le 1,000$ and $\| \bm{y} - P \bm{G}_{\bm{\theta}}\left( \bm{z} \right) \|_2^2 > 3^2 \cdot N \cdot H_{2D} \cdot W_{2D}$.

\begin{table}[htb]
  \begin{center}
    \caption{Means of metrics ($N=2$). We show variances in brackets.}
    \label{tab:results}    
    \begin{tabular}{lcc} \hline
          & standard-dose protocol & ultra low-dose protocol \\
      SSIM & 0.7675 (0.001931)& 0.7008 (0.0005670)  \\
      PSNR [dB]  & 25.89 (2.647)& 23.58 (0.6132)\\
      MAE & 0.02364 ($5.645 \times 10^{-5}$)& 0.02991 ($1.052 \times 10^{-5}$)  \\
      NRMSE  & 0.05731 (0.0002204) & 0.07349 ($5.007 \times 10^{-5}$) \\
      \hline 
    \end{tabular}
    \end{center}
\end{table}

For $N=2$, we show input 2D images without noise and 2D projections of 3D reconstructed images in Fig.~\ref{fig:inout}.
Moreover, we show a 3D chest CT image reconstructed from Figs.~\ref{fig:inout}(a) and (b) in Fig.~\ref{fig:typical_reconstructed_image_2_planes_wo_noise} and a differential image between the reconstructed 3D image and the ground-truth image in Fig.~\ref{fig:typical_diff_reconstructed_image_2_planes_wo_noise}.
We show enlarged axial and coronal slices in a pulmonary window setting in Fig.~\ref{fig:enlarge_lung}.
We show the means and variances of structural similarity (SSIM) \cite{wang2004image}, peak-signal-to-noise-ratio (PSNR), mean absolute error (MAE), and normalized root mean squared error (NRMSE) between the reconstructed 3D chest CT images and ground-truth images in Table~\ref{tab:results}.

\subsection{Ultra low-dose protocol}
\begin{figure}[t]
  \begin{minipage}[b]{0.45\hsize}
    \centering
    \includegraphics{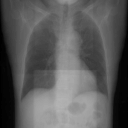}
    \subcaption{}\label{fig:pre1of2}
  \end{minipage}
  \begin{minipage}[b]{0.45\hsize}
    \centering
    \includegraphics{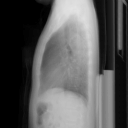}
    \subcaption{}\label{fig:pre2of2}
  \end{minipage}\\
  \begin{minipage}[b]{0.45\hsize}
    \centering
    \includegraphics{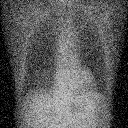}
    \subcaption{}\label{fig:noisy1of2}
  \end{minipage}
  \begin{minipage}[b]{0.45\hsize}
    \centering
    \includegraphics{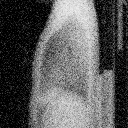}
    \subcaption{}\label{fig:noisy2of2}
  \end{minipage}\\
  \begin{minipage}[b]{0.45\hsize}
    \centering
    \includegraphics{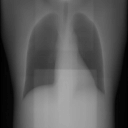}
    \subcaption{}\label{fig:33_noisy_recon1of2}
  \end{minipage}
  \begin{minipage}[b]{0.45\hsize}
    \centering
    \includegraphics{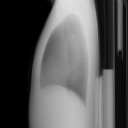}
    \subcaption{}\label{fig:33_noisy_recon2of2}
  \end{minipage}    
  \caption{Input images and projections of a reconstructed image ($N=2$): (a, b) projections of the ground-truth image, (c, d) noisy input 2D images assuming an ultra low-dose protocol, (e, f) projections of the optimum reconstructed image. The intensities of these images have been modified to enhance visibility.}
  \label{fig:inout_noisy}
\end{figure}

\begin{figure}
    \centering
    \includegraphics[width=8.5cm]{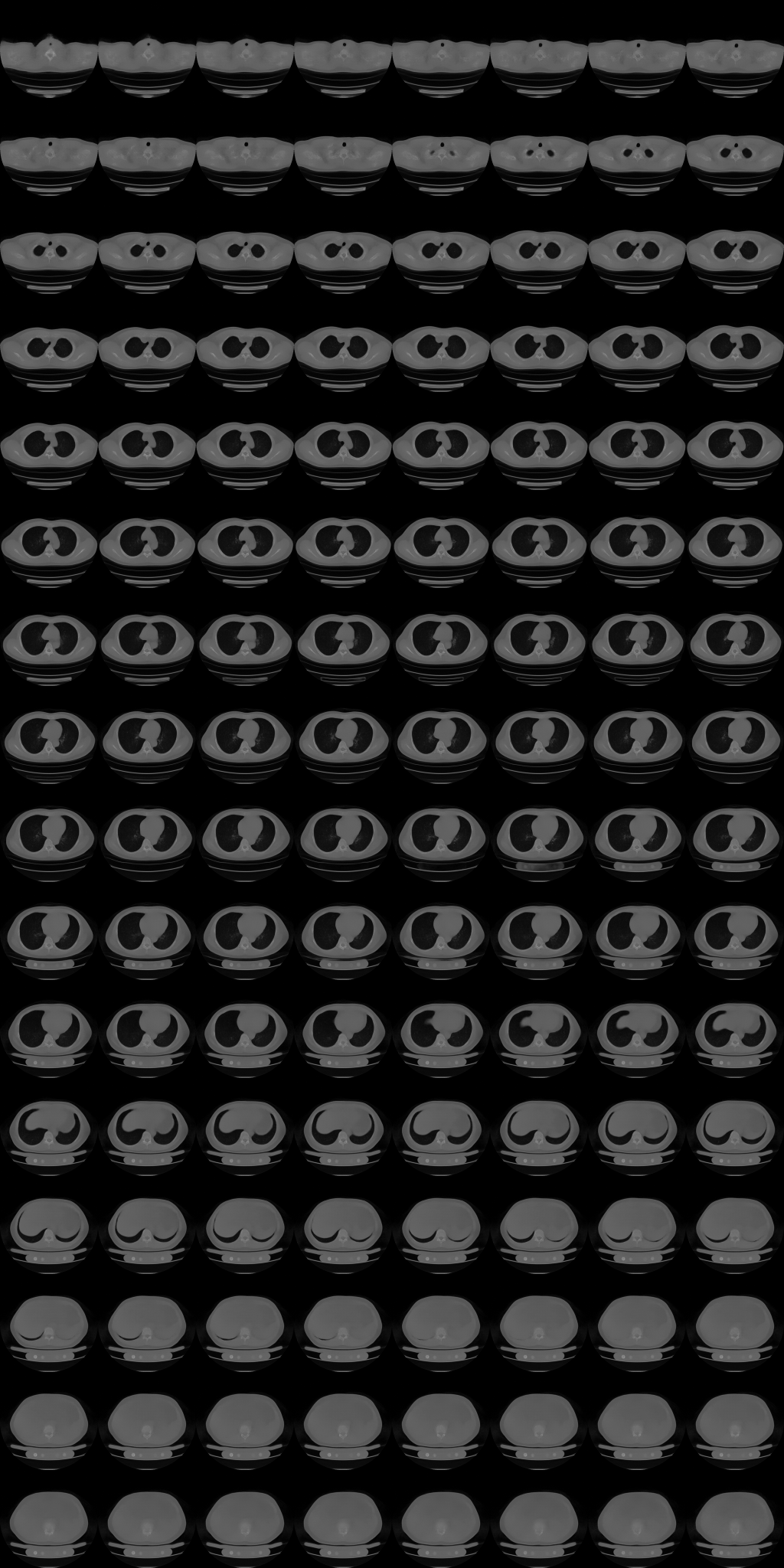}
    \caption{Reconstructed 3D CT image with X2CT-FLOW from Figs.~\ref{fig:inout_noisy}(c) and (d) ($\sigma^2=100, N=2$).}
    \label{fig:typical_reconstructed_image_2_planes_with_noise}
\end{figure}

\begin{figure}
    \centering
    \includegraphics[width=8.5cm]{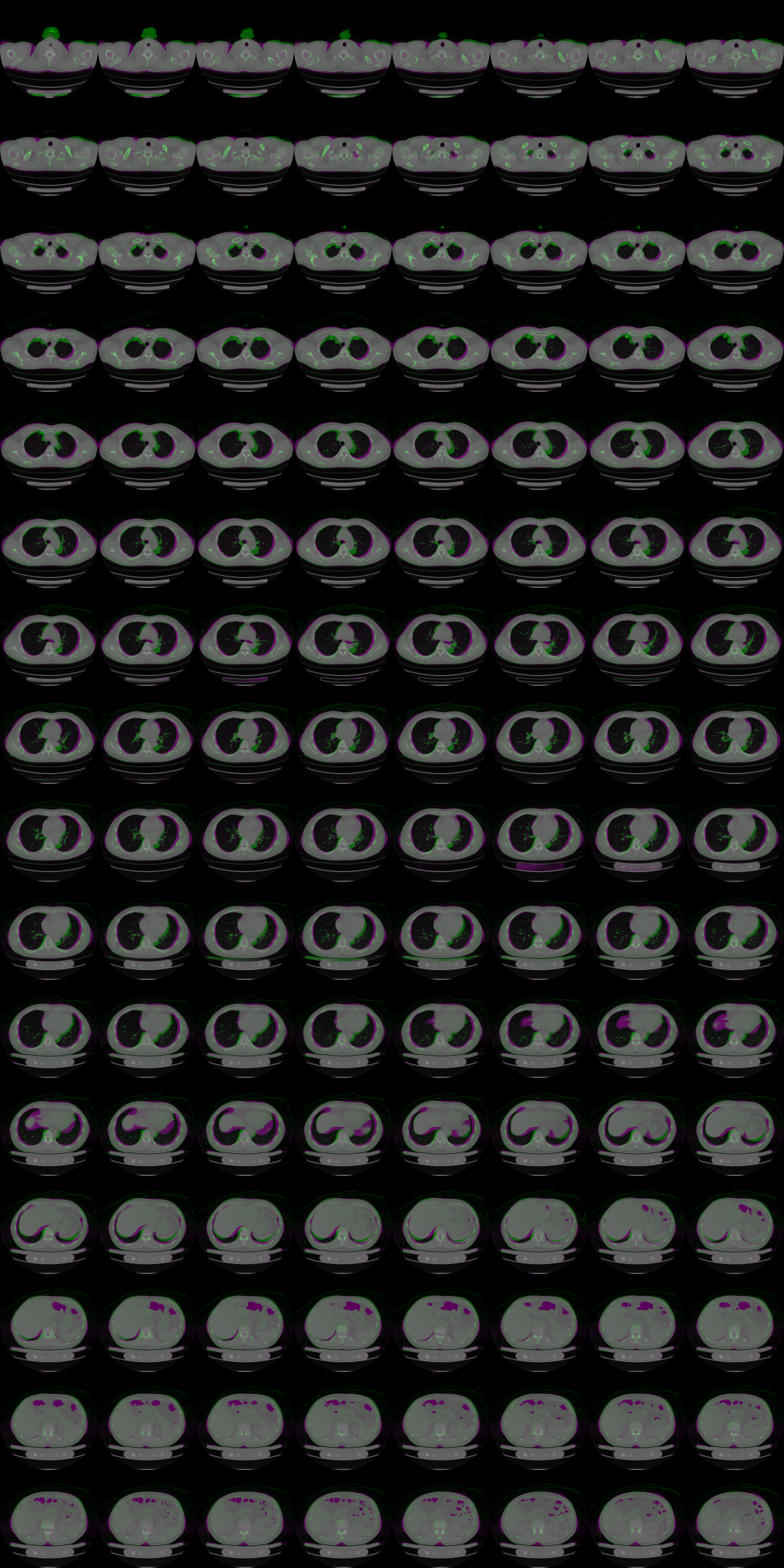}
    \caption{Superposition of the reconstructed 3D CT image shown in Fig.~$\ref{fig:typical_reconstructed_image_2_planes_with_noise}$ (magenta) and the ground-truth image (green).}
    \label{fig:typical_diff_reconstructed_image_2_planes_with_noise}
\end{figure}

To simulate an ultra low-dose protocol, we added an independent normal noise $\mathcal{N}(0, 10^2)$ to each 2D projection image $\bm{y}_i^j$.
We optimized Eq.~(\ref{eqn:bayes}) with $\sigma^2 = 100$ and $\alpha = 0.9 \cdot \left[ 1 - \exp{(-0.01 \cdot n)} \right]$.
We iterated while $n\le 1,000$ and $\| \bm{y} - P \bm{G}_{\bm{\theta}}\left( \bm{z} \right) \|_2^2 > 3^2 \cdot N \cdot H_{2D} \cdot W_{2D}$.
We empirically replaced the log-likelihood term $\log{p_{\bm{\theta}}(\bm{x})}$ with $\log{p_{\bm{\theta}}(\bm{x})^{T_b^2}}$, where $T^2_b=(\log{2} \cdot D_{3D} \cdot H_{3D} \cdot W_{3D})^{-1}$, i.e., bits per dimension.
For $N=2$, we show noisy input 2D images and 2D projection images of a 3D reconstructed image in Fig.~\ref{fig:inout_noisy}.
Moreover, we show a 3D chest CT image reconstructed from Figs.~\ref{fig:inout_noisy}(c) and (d) in Fig.~\ref{fig:typical_reconstructed_image_2_planes_with_noise}, and a differential image between the reconstructed 3D image and the ground-truth image in Fig.~\ref{fig:typical_diff_reconstructed_image_2_planes_with_noise}.
We show enlarged axial and coronal slices in a pulmonary window setting in Fig.~\ref{fig:enlarge_lung}.
We show the means and variances of SSIM, PSNR, MAE, and NRMSE between the reconstructed 3D chest CT images and ground-truth images in Table~\ref{tab:results}.

\section{Discussion}
\label{sec:discussion}
We designed X2CT-FLOW to find the optimum 3D chest CT image with MAP reconstruction.
We realized X2CT-FLOW by exploiting two features of the flow-based deep generative models: they can estimate the exact log-likelihood of an image, i.e., density estimation, and they can efficiently sample fictional but realistic images, i.e., sampling.
Unlike in related works \cite{kothari2021trumpets,asim2020invertible,whang2020compressed,whang2021composing,marinescu2020bayesian}, we reconstructed 3D CT images from 2D projection images.

In the limit of $\sigma^2 \rightarrow 0$, X2CT-FLOW finds 3D chest CT images whose projections onto each 2D plane are equivalent to each original input 2D projection image with the latent space exploration [Eq.~(\ref{eqn:gdm_z})].
Previous studies \cite{ying2019x2ct,peng2020xraysyn,shen2019patient,shen2021geometry} contain the cycle consistency loss for end-to-end supervised deep learning, but those losses are for training, hence, not for testing.
From this viewpoint, a related work is PULSE \cite{menon2020pulse}, but it deals with super-resolution between 2D images.
X2CT-FLOW deals with the reconstruction of optimum 3D chest CT images from a single or a few 2D projection images.

In the standard-dose protocol, while the initial guess images [Fig.~\ref{fig:inout}(c) and (d)] are clearly different from the input images [Figs.~\ref{fig:inout}(a) and (b)], the optimum reconstructed images [Figs.~\ref{fig:inout}(e) and (f)] well coincide with the input images.
Figures~\ref{fig:typical_reconstructed_image_2_planes_wo_noise} and \ref{fig:enlarge_lung} show that X2CT-FLOW can reconstruct the structure of organs (e.g., lungs, heart, and liver).
Moreover, X2CT-FLOW can well reconstruct the position of the bed.
However, X2CT-FLOW cannot well reconstruct finer structures, e.g., bronchovascular.
Therefore, this issue could affect the reconstruction of finer abnormalities.
This issue also could impact SSIM, PSNR, MAE, and NRMSE.

There are five possible extensions for X2CT-FLOW.
First, we emulated CT images in an ultra low-dose protocol using normal noise, but it is required to use authentic CT images in an ultra low-dose protocol to adopt X2CT-FLOW in clinical practice.
Second, we adopted the linear operator to take an average to obtain 2D projection images from a 3D chest CT image.
We can replace the linear operator with an arbitrary nonlinear differentiable operator from a 3D image to other images.
Moreover, we do not have to retrain the flow-based deep generative model when we change the operator.
Third, we limited the maximum number of projections for a 3D CT image to two planes ($N=2$), i.e., projections onto the sagittal and coronal planes.
However, it is possible to increase the number of projections if additional projection images are available.
This could contribute to enhancing SSIM, PSNR, MAE, and NRMSE, but it also enhances the radiation exposure.
Fourth, apart from 3D GLOW, our proposed method could be applied to other kinds of flow-based deep generative model, e.g., Flow++ \cite{ho2019flow++} and residual flows \cite{chen2019residual} if we extend those 2D models to 3D models.
Lastly, although we adopted the dataset of normal subjects, models trained with a dataset of abnormal subjects could be used to reconstruct 3D chest CT images with abnormalities.

Although we dealt with reconstruction of 3D chest CT images from clean or noisy 2D projection images, we can adopt the proposed algorithm to other applications apart from medical image analysis.
For example, we could apply X2CT-FLOW to estimate 3D shock wave structures from 2D Schlieren images, which are projection images of the air density gradient.

\section{Conclusions}
\label{sec:conclusions}
We proposed X2CT-FLOW built upon 3D GLOW for the MAP reconstruction of 3D chest CT images from a single or a few projection images.
To realize the practical high-resolution model, we newly developed the progressive learning.
We validated X2CT-FLOW by two numerical experiments assuming a standard-dose protocol or an ultra low-dose protocol.
The 3D chest CT images reconstructed from biplanar projection images without noise contamination showed good agreement with ground-truth images in terms of SSIM (0.7675 on average), PSNR (25.89 dB on average), MAE (0.02364 on average), and NRMSE (0.05731 on average).
Moreover, our images reconstructed from images contaminated with normal noise ($\mathcal{N}(0, 10^2)$) and the ground-truth images also showed good agreement in terms of SSIM (0.7008 on average), PSNR (23.58 dB on average), MAE (0.02991 on average), and NRMSE (0.07349 on average).
Further validations of X2CT-FLOW to adopt it for clinical practice are necessary, e.g., (i) validation for the reconstruction of abnormal lesions and (ii) validation using authentic CT images in an ultra low-dose protocol, which are included in our future works.

\appendices
\section{Codes for progressive learning}
\label{app:codes_progressive_learning}

\begin{lstlisting}[language=Python, label=code1]
import numpy as np
# src : source image (ndarray)
# n_bits_dst : the number of bits for the destination image
# (integer)
# n_bits_src : the number of bits for the source image
# (integer)
# return : destination image with reduction (ndarray)
def color_reduction(src, n_bits_dst = 4, n_bits_src = 8):
    dst = np.copy(src)
    delta = 2**n_bits_src // 2**n_bits_dst
    for c in range(2**n_bits_src // delta):
        inds = np.where((delta * c <= src) \
                               & (delta * (c + 1) > src))
        dst[inds] = (2 * delta * c + delta ) // 2
    return dst
\end{lstlisting}

\section{Ablation study for progressive learning}
\label{app:progressive_learning}

\begin{figure}
    \centering
    \includegraphics[width=8.5cm]{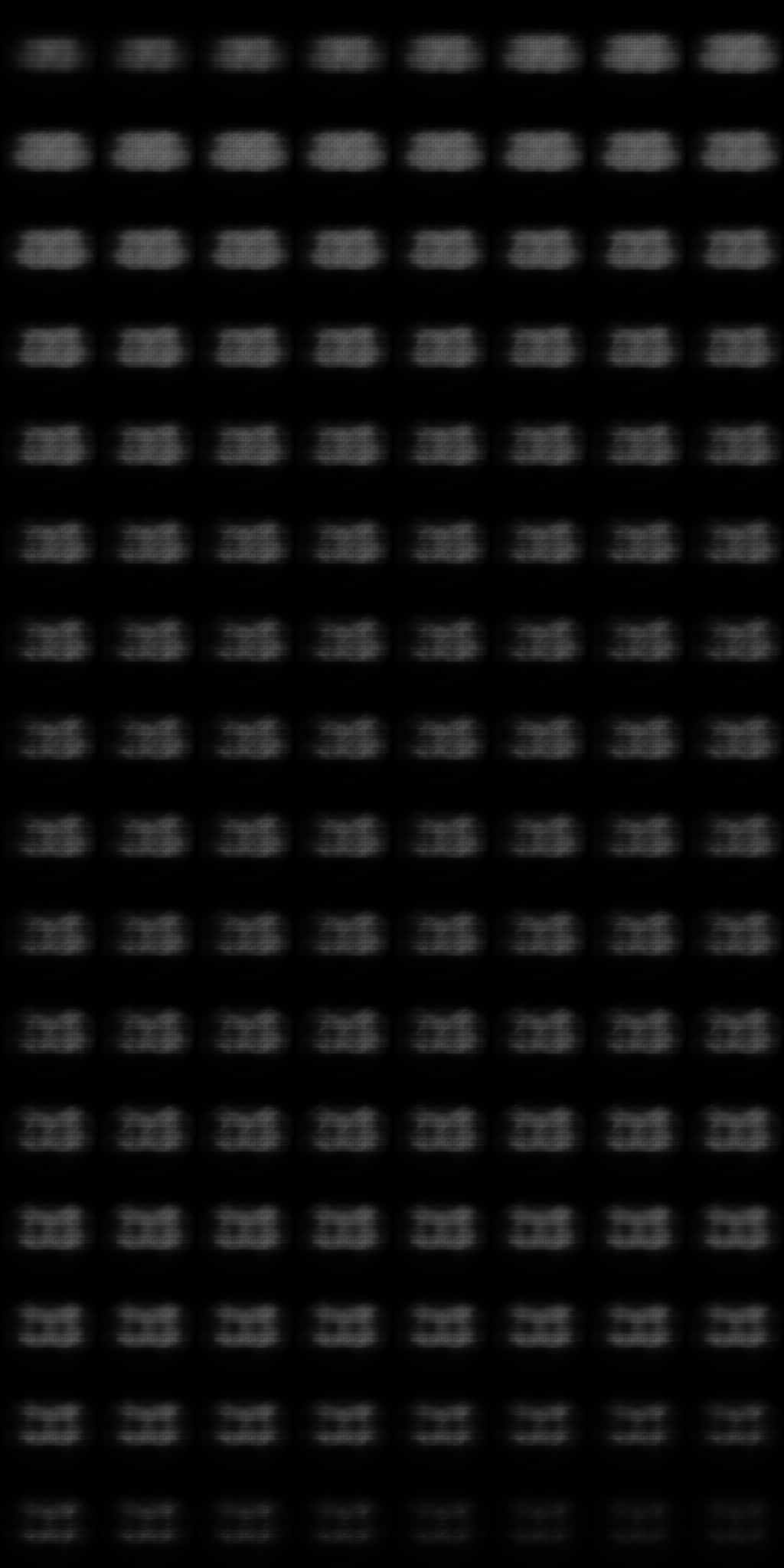}
    \caption{Fictional mean 3D CT image at 48 epochs (sampled with $T=0$, standard learning).}
    \label{fig:abrupt}
\end{figure}

\begin{figure}
    \centering
    \includegraphics[width=8.5cm]{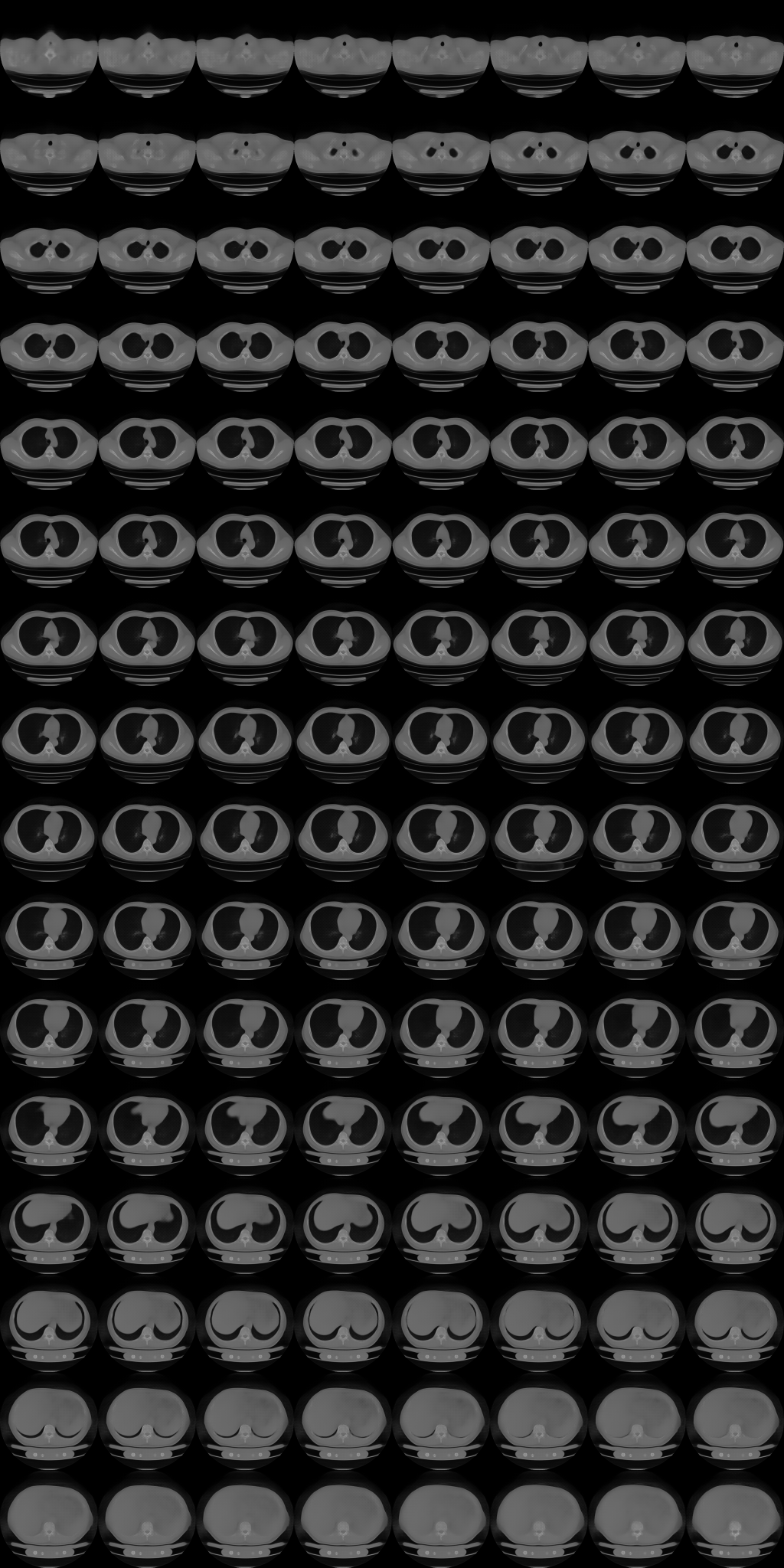}
    \caption{Fictional mean 3D CT image at the final epochs (sampled with $T=0$, progressive learning).}
    \label{fig:progressive}
\end{figure}

We abruptly started training the 3D chest CT model with 8 bits and continued it until 588 (= 96 + 324 + 24 + 144) epochs.
We validated the model once per 12 epochs. 
The validation loss took its minimum value at 48 epochs.
Figures~\ref{fig:abrupt} and \ref{fig:progressive} show sampling results with $T = 0$ for this standard learning (at 48 epochs) and the progressive learning (at the final epochs and when the model experienced the minimum loss in the 8 bits training), respectively.
These figures apparently show the superiority of the progressive learning.

\section{Formulations for $N\ge2$}
\label{app:formulation_n_gt_1}
We define other projection operators $P^j$, variances $(\sigma^j)^2$, projection images $\bm{y}^j$, and noise vector $\bm{w}^j$.
We distinguish projection directions by the superscript $j$.
We assume that there is no correlation among $\bm{w}^j$.
Therefore, we have
\begin{eqnarray}
    \bm{y}^j - P^j \bm{x} &=& \sqrt{{\left(\sigma^j\right)}^2} \bm{w}^j, \\
    \bm{w}^j &\sim& \mathcal{N}({0},\bm{I}). 
\end{eqnarray}

The log-posterior is now conditioned with all those projection images $\bm{y}^j$.
Therefore, we have
\begin{eqnarray}
 \label{eqn:bayes_multiple}
    \hat{\bm{x}} &=& \argmax_{\bm{x}} \log p (\bm{x} |  \bm{y}^1, \bm{y}^2, \ldots, \bm{y}^N) \nonumber \\
    &=& \argmax_{\bm{x}} \log p (\bm{y}^1, \bm{y}^2, \ldots, \bm{y}^N| \bm{x}) + \log p(\bm{x}) \nonumber\\
    &=& \argmax_{\bm{x}} \sum_j \log p (\bm{y}^j | \bm{x})  + \log p(\bm{x}) \nonumber\\
    &=&  \argmax_{\bm{x}} \sum_j \log \left[ \frac{1}{\sqrt{2\pi {\left(\sigma^j\right)}^2}}\exp{\left(-\frac{1}{2} \left(\bm{w}^j\right)^T \bm{w}^j \right)}  \right] \nonumber\\
    &+& \log p(\bm{x})  \nonumber \\
    &=& \argmax_{\bm{x}} \sum_j -\frac{1}{2 {\left(\sigma^j\right)}^2} \| \bm{y}^j - P^j \bm{x} \|_2^2 \nonumber \\  
    &+& \log p(\bm{x})  \\
    &\equiv& \argmax_{\bm{x}} - \mathcal{E'}\left(\bm{x}\right).
\end{eqnarray}
In the deformation from the second line to the third line,
we applied the fact that normal noise distributions among $\bm{y}^j$ are independent of each other.

\bibliographystyle{IEEEtran}
\bibliography{main.bib}

\end{document}